\documentclass[
superscriptaddress,
preprint,
 amsmath,amssymb,
 aip,
]{revtex4-2}

\usepackage{graphicx}
\usepackage{dcolumn}
\usepackage{bm}

\usepackage{color}

\begin{document}

\title{Ultrafast melting of charge-density wave fluctuations at room temperature in $\mathbf{1T\mathchar`-TiSe_2}$ monitored under non-equilibrium conditions
}

\author{Yu Mizukoshi}
\email{s2220292@s.tsukuba.ac.jp}
\affiliation{Department of Applied Physics, Graduate school of Pure and Applied Sciences, University of Tsukuba, 1-1-1 Tennodai, Tsukuba 305-8573, Japan}
\author{Takumi Fukuda}
\affiliation{Department of Applied Physics, Graduate school of Pure and Applied Sciences, University of Tsukuba, 1-1-1 Tennodai, Tsukuba 305-8573, Japan}
\author{Yuta Komori}
\affiliation{Department of Applied Physics, Graduate school of Pure and Applied Sciences, University of Tsukuba, 1-1-1 Tennodai, Tsukuba 305-8573, Japan}
\author{Ryo Ishikawa}
\affiliation{Graduate School of Science and Engineering, Saitama University, Saitama 338-8570, Japan}
\author{Keiji Ueno}
\affiliation{Graduate School of Science and Engineering, Saitama University, Saitama 338-8570, Japan}
\author{Muneaki Hase}
\email{Authors to whom correspondence should be addressed: mhase@bk.tsukuba.ac.jp}
\affiliation{Department of Applied Physics, Graduate school of Pure and Applied Sciences, University of Tsukuba, 1-1-1 Tennodai, Tsukuba 305-8573, Japan}

\date{\today}

\begin{abstract}
We investigate the ultrafast lattice dynamics in $\mathrm{1T\mathchar`-TiSe_2}$ using femtosecond reflection pump-probe and pump-pump-probe techniques at room temperature. The time-domain signals and Fourier-transformed spectra show the $A_{1g}$ phonon mode at 5.9 THz. Moreover, we observe an additional mode at $\approx$ 3 THz, corresponding to the charge-density wave (CDW) amplitude mode, which is generally visible below T$_c \approx 200\ $K. We argue that the emergence of the CDW amplitude mode at room temperature can be a consequence of fluctuations of order parameters, based on the additional experiment using the pump-pump-probe technique, which exhibited suppression of the AM signal within the ultrafast time scale of $\sim$ 0.5 ps.
\end{abstract}

\maketitle

\newpage

Two-dimensional (2D) materials, such as graphene and transition-metal dichalcogenides (TMDCs), exhibit a variety of exotic quantum states owing to the strong electronic correlations by the quantum confinement effects within each atomic layer. 
For example, 2D materials give rise to the large binding or higher-order intra- and interlayer exciton,\cite{Chernikov2014Exciton, you2015observation, Arora2017Interlayerexciton} unconventional charge-density-wave (CDW) phases,\cite{xi2015strongly,kogar2017signatures} and topological phases depending on the lattice inversion symmetry that can be manipulated by the chemical bonding or stacking orders.\cite{Cho2015Phasepatterning, chen2016activation}
Importantly, ultrafast photo-excitation of 2D materials can be a tool for tuning the properties of exotic quantum phases of matters under non-equilibrium conditions, such as electron-hole correlations,\cite{chernikov2015population, sie2017observation} manipulations of the CDW orders,\cite{schmitt2008transient,eichberger2010snapshots,kogar2020light,zong2021role} and lattice symmetry switching of topological phases.\cite{sie2019ultrafast, Zhang2019LILSS, Fukuda2020APL, cheng2022Persistent,Han2022JPCL}
These non-equilibrium phenomena are associated with the electron-hole or electron-lattice interactions, which significantly modify the macroscopic electronic properties, such as electron mobility. Thus, exploring the ultrafast response in 2D materials is essential to gain insight into the possibility of developing next-generation device applications under ultrafast control.

\begin{figure}[p]
    \centering
    \includegraphics[width = 8.8cm]{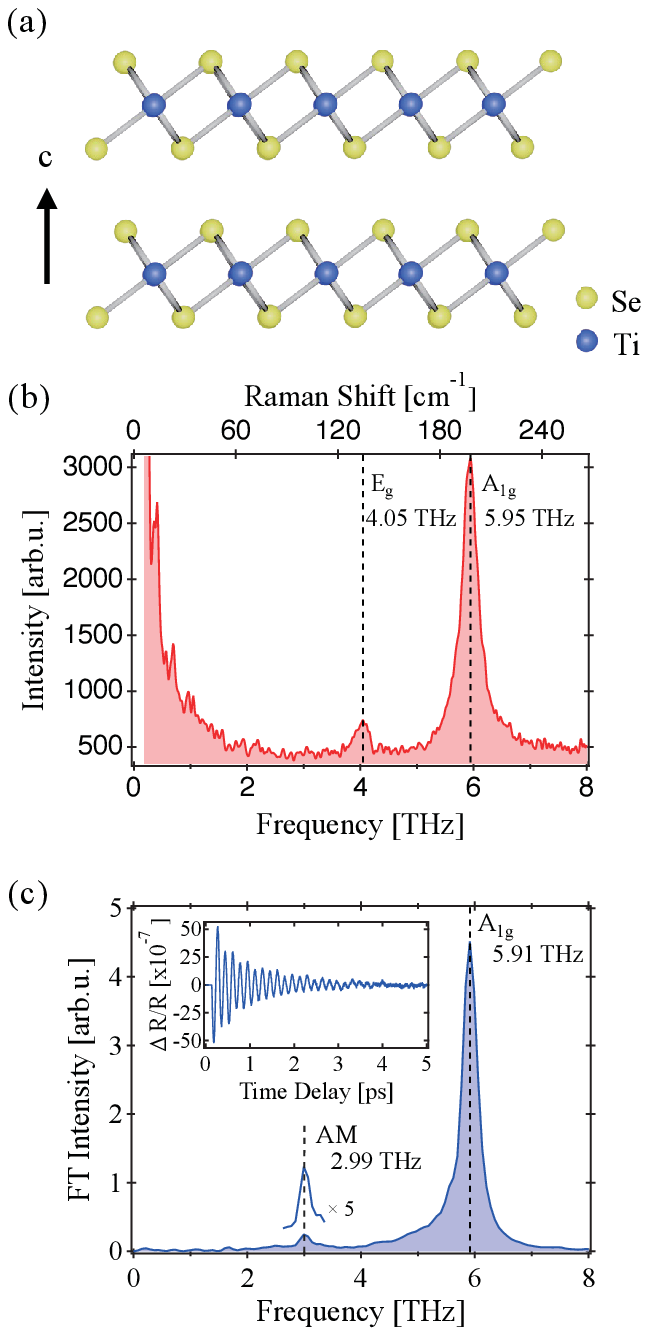}
    \caption{
    (a) Schematic crystal structure of 1T-TiSe$_2$. 
    (b) Raman scattering spectra at room temperature. The E$_g$ and A$_{1g}$ optical modes are observed at 4.05 THz and 5.95 THz, respectively. 
    (c) FT spectra of the time domain QP signals shown in the inset for the pump fluence of 70 $\mu$J/cm$^2$.
    Two different peaks are observed, originating from the AM at 2.99 THz and $A_{1g}$ mode at 5.91 THz.
    The inset shows the time-domain signals of the QP oscillations extracted from $\Delta R / R$ data.
    }
    \label{figure1}
\end{figure}

$\mathrm{1T\mathchar`-TiSe_2}$, a family of TMDCs with a Ti atom sandwiched between two layers of Se atoms, has attracted considerable attention due to the presence of a unique CDW phase transition.
The crystal structure of $\mathrm{1T\mathchar`-TiSe_2}$ is shown in Fig. \ref{figure1}(a). At room temperature, $\mathrm{1T\mathchar`-TiSe_2}$ is a normal semimetal phase.
Below $T_c \approx 200\ \mathrm{K}$, it undergoes a second-order phase transition into a commensurate CDW-ordered phase with a $(2\times 2\times 2)$ periodic lattice distortion. \cite{holy1977raman}
The mechanism for the CDW phase transitions has been debated for decades and remains elusive. \cite{kidd2002electron,porer2014non,mathias2016self,kogar2017signatures}
Recent interests in $\mathrm{1T\mathchar`-TiSe_2}$ have rather focused on the phase transition phenomena above $T_c$.
Despite $T > T_c$, several studies found the competition or coexistence between CDW disordered and ordered states, which are considered to originate from the fluctuations of order parameters around $T_c$.\cite{holt2001x,cercellier2007evidence,monney2010temperature,monney2016revealing,chen2016hidden}
In addition to the clear observations of CDW fluctuations persisting up to 250 K, \cite{holt2001x,cercellier2007evidence,monney2010temperature,monney2016revealing,chen2016hidden} a few studies suggest that they survive even at room temperature.\cite{zhang2022second,cheng2022light}
Although, short-range CDW state can be created by defect doping for $T > T_c$,\cite{sahoo2022short} CDW fluctuations are dynamic phenomena, i.e., fluctuate in the time and spatial domains, whose behavior at room temperature has not been fully elucidated.

Here, we investigate the ultrafast dynamics of CDW fluctuations in the $\mathrm{1T\mathchar`-TiSe_2}$ bulk crystal using reflection-type femtosecond optical pump-probe and pump-pump-probe techniques under ambient conditions.
In the pump-probe measurement, we found CDW-induced collective mode, i.e., the CDW amplitude mode (AM), which cannot be observed by Raman scattering.
The pump-pump-probe measurements further reveal that the photo-excitation suppresses the AM signal within the ultrafast time scale of $\sim$ 0.5 ps.
Our results suggest that CDW fluctuations survive up to room temperature under equilibrium conditions and quench under photo-excited non-equilibrium conditions.
Observation and quenching of the CDW fluctuations will open a route to ultrafast switching of the local CDW states at room temperature for possible device applications based on CDW current.\cite{Mohammadzadeh2021}

The sample used was a bulk single crystal of $\mathrm{1T\mathchar`-TiSe_2}$ grown by flux zone method (commercially available from \textit{2D\ Semiconductors}, USA).
The sample purity was 99.9999 \% with defect concentration as low as $10^9 \sim 10^{10}$ cm$^{-2}$.
Since our sample is impurity-free single crystal, the defect-induced CDW fluctuations cannot be applicable.\cite{sahoo2022short}
The crystal thickness used in this study was $\sim$ 100 $\mu$m.
The conventional Raman scattering measurements were conducted using InVia (Renishow) under 532 nm laser excitation with the power of 1.5 mW at room temperature (295 K) under ambient conditions.

The optical pump-probe and pump-pump-probe measurements were performed at room temperature using a Ti:sapphire oscillator (Element 2, Spectra-Physics) with a central wavelength of 795 nm (photon energy of 1.56 eV), a pulse width of $\leq$ 10 fs, and a repetition rate of 75 MHz. In the pump-pump-probe experiment, the pump beam was split into two beams by a Michelson interferometer. The $s$-polarized pump (and pre-pump) and the $p$-polarized probe beam were focused onto the sample to a spot diameter of $\sim 20\ \mathrm{\mu m}$.
The transient isotropic reflectivity changes $\Delta R / R$ were measured as a function of the pump-probe time delay by scanning the optical path length of the probe beam at 9.5 Hz.
By removing the photo-excited carrier response from the $\Delta R / R$ signal by the fit with an exponential decay function, the coherent oscillation parts of the $\Delta R / R$ signal were extracted.
All measurements were performed under ambient conditions at room temperature (295 K).

Figure \ref{figure1}(b) shows a Raman scattering spectrum, indicating the presence of the $E_g$ and  $A_{1g}$ optical modes at 4.05 THz and 5.95 THz, respectively. 
The obtained Raman spectrum is consistent with the previous studies.\cite{holy1977raman,SUGAI1980433,duong2017raman}
The inset of Fig. \ref{figure1}(c) shows coherent oscillation parts of transient reflectivity change $\Delta R / R$ measured at the pump fluence of 70 $\mu$J/cm$^{2}$, which were clearly observed up to $\sim$ 4 ps.
These coherent oscillations arise from photo-generated quasi-particles (QPs): i.e., coherent phonons, traditionally described by the resonant impulsive stimulated Raman scattering (ISRS) or by the displacive excitation of coherent phonons (DECPs) generation mechanisms.\cite{zeiger1992theory}
The Fourier transformed (FT) spectrum of the QP signals is shown in Fig. \ref{figure1}(c), where one can see a strong peak observed at 5.91 THz that is in good agreement with the $A_{1g}$ optical mode \cite{holy1977raman,duong2017raman} [See also the Raman spectrum in Fig. 1(b)].
The anisotropic $E_g$ mode at 4.05 THz is not observed in the FT spectrum, possibly due to less sensitivity of anisotropic modes under the isotropic $\Delta R / R$ measurements. \cite{hase1996optical}
Interestingly, another peak located at 2.99 THz was found, which was not observed in the Raman spectrum [Figure \ref{figure1}(b)]. The peak is identical to the frequency of the CDW-induced collective mode, namely the amplitude mode (AM).
Compared to the previous Raman study, the frequency of 2.99 THz shows a red-shift from the literature value of 3.48 THz observed at 53 K. \cite{holy1977raman}
A red-shift was also reported during the melting of CDW long-range order under the irradiation by ultrafast light pulses below $T_c$, \cite{mohr2011nonthermal,hedayat2021investigation} i.e., during the coexistence regime of CDW and semimetal phases. In that case, the frequency observed in the present study is in good agreement with the literature value of 2.95 THz at 80 K. \cite{hedayat2021investigation}
Thus, the peak at 2.99 THz can be considered to be the AM, implying the existence of the CDW state, although the lattice temperature is above $T_c$.
Note that oxidation or thermal damage of the sample does not provide QPs around 2.99 THz. \cite{duong2017raman,lioi2017photon}

The presence of the AM in the pump-probe measurement is in contrast to the absence of the AM in static Raman measurement [see Fig. \ref{figure1}(b) and Ref.\cite{holy1977raman,duong2017raman}].
To explain the appearance of the AM in the pump-probe measurement, we first consider a hypothesis that the formation of a transient CDW-order under ultrafast photoexcitation in a way similar to the case of the rare-earth tritellurides family.\cite{kogar2020light,zong2021role}
However, this is not the case for 1T-TiSe${_2}$ above $T_c$: a previous experimental report did not show a transient CDW-order in the form of emergent superlattice diffraction peaks upon photoexcitation in the presence of strong CDW fluctuations.\cite{cheng2022light}
Furthermore, in general, the scattering of carriers and phonons gives rise to quenching of the CDW state.\cite{mohr2011nonthermal,porer2014non,mathias2016self,monney2016revealing}
Therefore, a possible hypothesis is that CDW fluctuations play a central role in the emergence of the AM oscillation.
Note that CDW fluctuations are the phenomena that the CDW state persists even above $T_c$ as like fluctuations of order parameters with short-range coherence, which have been occasionally observed and discussed in several kinds of literature using X-ray and electron diffraction measurements,\cite{holt2001x, cheng2022light} angle-resolved photoemission spectroscopy,\cite{cercellier2007evidence,monney2010temperature,monney2016revealing,chen2016hidden} and second-harmonic generation spectroscopy.\cite{zhang2022second}
The presence of the AM in our pump-probe experiment indicates that the CDW fluctuations at room temperature can be observed preferably by pump-probe techniques but would be difficult to see by static measurements, such as Raman scattering. Although the reason for this is not yet fully understood, a plausible reason for this is the fact that the CDW fluctuations are the time-dependent phenomena and will be time averaged when using static Raman measurement. Under the simple pump-probe experiment, the pump pulse induces coherent CDW fluctuations, which can be observed by the delayed probe pulse.

\begin{figure}[p]
    \centering
    \includegraphics[width = 10.8cm]{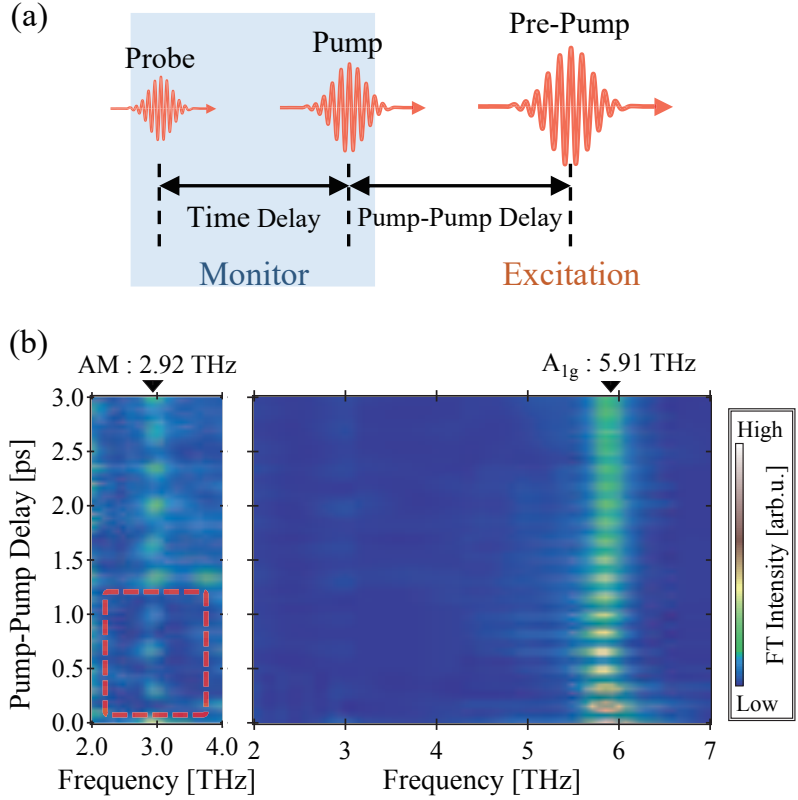}
    \caption{(a) Schematic of the pump-pump-probe measurement. 
    The pump-pump delay indicates the time interval between the pre-pump and pump pulses, whereas the time delay indicates the time interval between the pump and probe pulses.
    (b) Two-dimensional spectra of FT intensity as the functions of the pump-pump delay and frequency. The left panel focuses on the AM and the right panel focuses on the A$_{1g}$ mode for different frequency regions.
    Each mode shows a periodic modulation, which is consistent with each frequency.
    The dashed red rectangle on the left panel indicates the region where the AM intensity was weakened by the pre-pump pulse irradiation.
    }
    \label{figure2}
\end{figure}

To further examine the possible mechanism of CDW emergence, we performed the pump-pump-probe measurement,\cite{yusupov2010coherent,KOSHIHARA20221,Yu2023manipulation} which clearly exhibits the dynamics of evolution or suppression of the CDW order, and the subsequent recovery of the CDW order as shown below. 
In this measurement, the pump pulse was split into two beams;
the initial stronger pre-pump pulse ($\approx$ 140 $\mu$J/cm$^{2}$) 
brings the initial state into a photo-induced non-equilibrium state, after that the second pump pulse ($\approx$ 70 $\mu$J/cm$^{2}$) 
generates QPs, and then, the probe pulse monitors the transient reflectivity change $\Delta R / R$, as schematically shown in Fig. \ref{figure2}(a).
The fluence of the pre-pump pulse was chosen to the twice that of the second pump pulse, so as to create the high-density photoexcited state.\cite{Hase2015GST}
The QP signals indicate a transient state under the photoexcited non-equilibrium condition, extracted after the second pump pulse for various pump-pump time delays.
The pump-pump delay ($\Delta t$) was varied in steps of $\sim$ 40 fs, which was sufficiently shorter than the QPs phonon periods.
We have followed the approach of Ref. \cite{yusupov2010coherent}, in which the pump-pump separation time ($\Delta t$) did not match with the phonon period.

Figure \ref{figure2}(b) shows the time-frequency chronogram as a function of the frequency and pump-pump delay.
The two QP signals are observed at the frequency of 2.92 THz for the AM and 5.91 THz for the $A_{1g}$ phonon mode.
The periodic modulation of the FT intensity was observed for the AM and $A_{1g}$ phonon modes, whose periodicity coincides with each QP frequency,
as confirmed by comparing the periodic modulation with the cosine oscillation with the frequencies of 2.92 and 5.91 THz, respectively (data not shown).
The coherent modulations have been known as coherent control of QPs, resulted from the coherent superposition of QPs or could be understood as the interference of coherent oscillations generated by the pre-pump and second-pump pulses.\cite{hase1996optical}
The interference signature of the AM was weakened at the short pump-pump delays, while the signature of the $A_{1g}$ mode simply decreased with the pump-pump delay [see the left side of Fig. \ref{figure2}(b)].
The difference observed for the two modes will be a hint for the emergence of the AM above $T_c$.

\begin{figure}[p]
    \centering
    \includegraphics[width = 16cm]{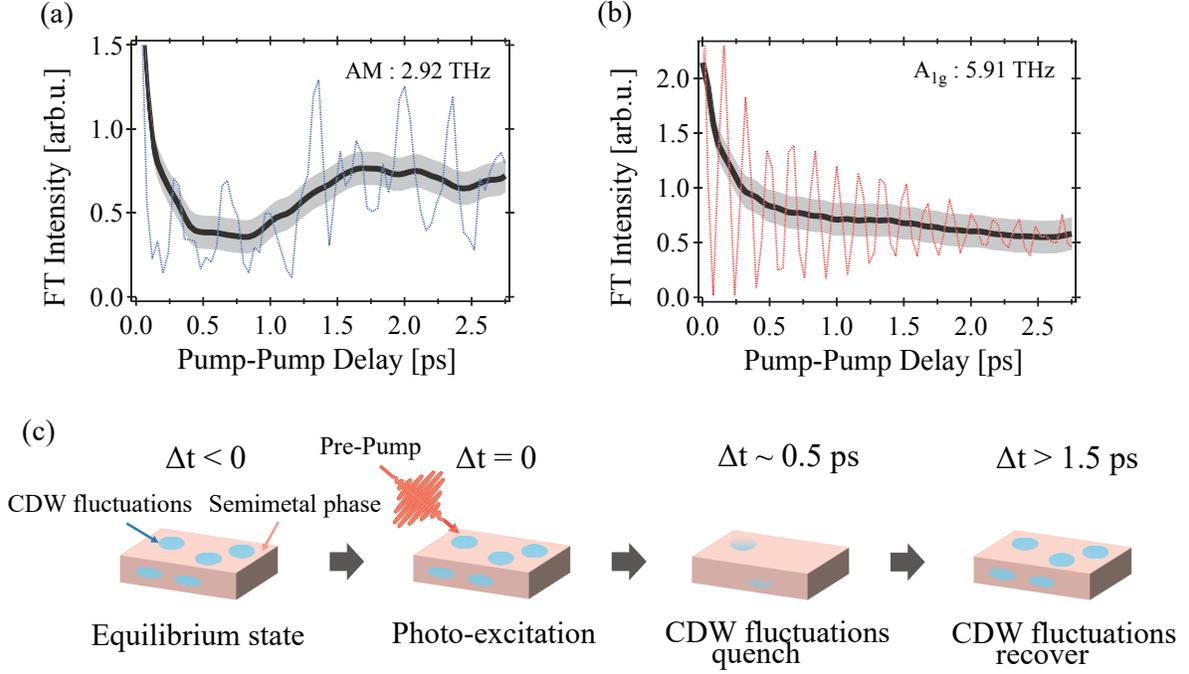}
    \caption{(a) FT intensity of the AM as a function of the pump-pump delay ($\Delta t$). The black thick line indicates the average of the FT intensity for two cycles. 
    (b) Same as (a) but for the A$_{1g}$ mode. Only for the AM, FT intensity depletes between 0.25 and 1.5 ps.
    (c) Schematics of the dynamics of CDW fluctuations. The light red and blue domains represent the semimetal phase and CDW fluctuations, respectively. 
    The CDW fluctuations partially exist in the dominant semimetal phase under an equilibrium state at room temperature ($\Delta t <$ 0).
    When the pre-pump pulse is irradiated ($\Delta t = 0$), the CDW fluctuations are partially quenched ($\Delta t \sim$ 0.5 ps).
    After that, the CDW fluctuations recover ($\Delta t >$ 1.5 ps).
    }
    \label{figure3}
\end{figure}

To explore the dynamics of the CDW state, we are most interested in  the FT intensity of each QP, plotted as a function of the pump-pump delay in Fig. \ref{figure3}(a) and \ref{figure3}(b).
Here, to eliminate the coherent modulations, the non-oscillation part of the FT intensity was extracted by averaging the coherent modulations for the two periods for each QP [see black thick line in Fig. \ref{figure3}(a), (b)].
From the averaged intensities obtained, suppression of the AM has been uncovered within $\sim$ 1.5 ps, while the $A_{1g}$ phonon shows an expected monotonic dephasing induced by the pre-pump pulse. 
The AM in Fig. \ref{figure3}(a) for $\Delta t< 0.25\ \mathrm{ps}$ shows a much stronger signal due to the constructive generation of coherent CDW fluctuations. The suppression (or quenching) of the AM is delayed by $\Delta t \sim$ 0.5 ps, suggesting that the CDW state melts instantaneously and then it recovers after $\sim$ 1.5 ps.
The time scale of $\Delta t \sim 0.5\ \mathrm{ps}$ implies that the quenching of CDW fluctuations is caused by electronic effect rather than lattice heating.
In fact, according to the two-temperature model (TTM), the first pre-pump pulse increases the electron temperature of $\Delta T_e \approx$ 1000 K, then subsequently electron-phonon coupling induces the elevating the lattice temperature of $\Delta T_l \approx 10\ \mathrm{K}$ after 1.5 ps.
Thus, the increased $T_e$ would destroy the correlations in the coherent CDW fluctuations either by topological defect \cite{Nelson1979Defect,Zong2019defect} or conventional CDW melting.\cite{mohr2011nonthermal}
Then, $T_e$ is cooled down due to electron-phonon interaction after $\Delta t > 1.5\ \mathrm{ps}$, and the CDW state will recover.
In addition, the time scale of quenching of CDW fluctuations ($\sim$ 0.5 ps) does not contradict with that of ultrafast melting of CDW fluctuations ($\sim$ 0.2--1.0 ps) observed by time-resolved electron diffraction at 250 K.\cite{cheng2022light} 
Therefore, in the present study, CDW fluctuations will be the more plausible explanation for the emergence of the AM oscillation than the light-induced creation effect.
It is noted that the existence of the coherent interference signature of the AM [blue dashed line in Fig. \ref{figure3}(a)] indicates the CDW fluctuations were not completely suppressed by the pre-pump pulse, but rather partially suppressed.

Based on the discussion above, we present the scheme of the interpretation of our observations in Fig. \ref{figure3}(c).
Under an equilibrium state at room temperature, the CDW fluctuations survive in the dominant semimetal phase ($\Delta t < 0$).\cite{chen2016hidden,zhang2022second,cheng2022light}
When the pre-pump pulse is irradiated ($\Delta t = 0$), the CDW fluctuations are partially suppressed (melt) due to the electronic effect within the ultrafast time scale of $\Delta t \sim$ 0.5 ps. Then the CDW fluctuations recover for $\Delta t >$ 1.5 ps.
Although these processes can be observed on the pump-pump-probe measurement under non-equilibrium conditions, they cannot be visible on the single-pump-probe measurement. 
On the single-pump-probe measurement, we will observe coherent QP signals induced only at $\Delta t \sim$ 0 ps; in other words, CDW fluctuations partially survived after photo-excitation. In contrast, on the pump-pump-probe measurement, we observed the snapshots of the dynamical suppression (melt) and recovery as the function of the pump-pump delay ($\Delta t$).  

To summarize, we explored ultrafast dynamics of CDW fluctuations in a prototypical TMDC, 1T-TiSe$_2$ bulk crystal, at room temperature much higher than $T_c$ using reflection-type femtosecond pump-probe and pump-pump-probe techniques. 
The pump-probe measurement uncovered CDW-induced collective mode, i.e., the CDW amplitude mode, which could not be observed by Raman scattering.
The pump-pump-probe measurements further revealed that the photo-excitation by the stronger pre-pump pulse suppresses the AM signal within the ultrafast time scale of $\sim$ 0.5 ps, while the $A_{1g}$ phonon shows a simple exponential decay.
The ultrafast time scale of $\sim$ 0.5 ps for the suppression (or quenching) of the AM suggests that the CDW state melts instantaneously by the electronic effect rather than lattice heating and then it recovers after $\sim$ 1.5 ps due to electron-phonon interaction.
The ultrafast dynamics of charge-density wave fluctuations above the transition temperature observed in the sub-picosecond time scale is expected to be a key phenomenon in controlling the quantum phases of matter and allow the 
development of ultrafast phase switching device applications, such as data storage and processing at room temperature.\cite{Mihailovic2002,Balandin2021}

\section*{Acknowledgement}
This work was supported by JSPS Research Fellowships for Young Scientists, JSPS KAKENHI (Grant Numbers. 19H02619, 21K04826, 21J20332, 22H01151, and 22H05445) and CREST, JST (Grant Number. JPMJCR1875).

\section*{AUTHOR DECLARATIONS}
\section*{Conflict of Interest}
The authors have no conflicts to disclose.

\section*{DATA AVAILABILITY}
The data that support the findings of this study are available from the corresponding author upon reasonable request.

\bibliography{APLmizukoshi}

\newpage

\end{document}